\documentclass[conference]{IEEEtran}
\IEEEoverridecommandlockouts
\usepackage{cite}
\usepackage{amsmath,amssymb,amsfonts}
\usepackage{algorithmic}
\usepackage{graphicx}
\usepackage{textcomp}
\usepackage{xcolor}
\usepackage{algorithm}
\usepackage{algorithmic} 
\usepackage{subfigure}
\usepackage{booktabs}
\usepackage{bbding}
\usepackage{caption}

\def\BibTeX{{\rm B\kern-.05em{\sc i\kern-.025em b}\kern-.08em
    T\kern-.1667em\lower.7ex\hbox{E}\kern-.125emX}}
\begin{document}

\title{Deep Joint Source-Channel Coding for Wireless Image Transmission with Semantic Importance}
\author{  Qizheng Sun,
  	Caili Guo,
  	Yang Yang,
  	Jiujiu Chen,
  	Rui Tang,
  	Chuanhong Liu\\
\IEEEauthorblockA{
  		Beijing University of Posts and Telecommunications,\\
  		Beijing Key Laboratory of Network System Architecture and Convergence,\\
  		Beijing Laboratory of Advanced Information Networks,
  		 100876, Beijing, China\\
  		Email: \{qizheng\_sun, guocaili, yangyang01, chenjiujiu, 1361680482, 2016\_liuchuanhong\}@bupt.edu.cn}
  	\thanks{This work is supported by the Beijing Natural Science Foundation (No.4202049); the Fundamental Research Funds for the Central Universities (2021XD-A01-1); BUPT Excellent Ph.D.Students Foundation (CX2021101).}
  	}

\maketitle

\begin{abstract}

The sixth-generation mobile communication system proposes the vision of smart interconnection of everything, which requires accomplishing communication tasks while ensuring the performance of intelligent tasks. A joint source-channel coding method based on semantic importance is proposed, which aims at preserving semantic information during wireless image transmission and thereby boosting the performance of intelligent tasks for images at the receiver. Specifically, we first propose semantic importance weight calculation method, which is based on the gradient of intelligent task’s perception results with respect to the features. Then, we design the semantic loss function in the way of using semantic weights to weight the features. Finally, we train the deep joint source-channel coding network using the semantic loss function. Experiment results demonstrate that the proposed method achieves up to 57.7\% and 9.1\% improvement in terms of intelligent task’s performance compared with the source-channel separation coding method and the deep source-channel joint coding method without considering semantics at the same compression rate and signal-to-noise ratio, respectively.

\end{abstract}


\section{Introduction}
Modern communication systems employ the separate source-channel coding method (SSCC) for the transmission of image data. SSCC first use the source coding algorithm (such as JPEG, WebP, BPG) to compress the image, and then use the source-independent channel coding method (such as LDPC, Polar, Turbo, etc.) to encode the bit stream, which is completed under the guidance of Shannon's separation theorem \cite{shannon2001mathematical}. This theorem shows that when transmitting infinite bits, it is optimal to split the communication task into (i) removing the redundant information of the source as much as possible and (ii) re-introducing redundant information for message reconstruction in the presence of channel noise.
However, in practice, a finite number of bits is transmitted, which can not satisfy infinite bits' assumption of Shannon's separation theorem. In fact, the signal quality at the receiver is affected by the joint influence of source coding distortion and channel coding error \cite{yang2021deep}, so it is necessary to jointly consider source coding and channel coding. In addition, the rapid development of deep learning technology enables the joint consideration of source coding and channel coding. Deep learning-based joint source-channel coding (deep JSCC) refers to use deep neural networks to implement source encoding and channel encoding by end-to-end (E2E) semantic communication framework. Deep JSCC uses the powerful learning ability of the neural network to learn how to remove the redundancy of source information and how to resist the channel noise \cite{yang2021deep}.
\begin{figure}[t]
	\centering
	\includegraphics[width=0.7\linewidth]{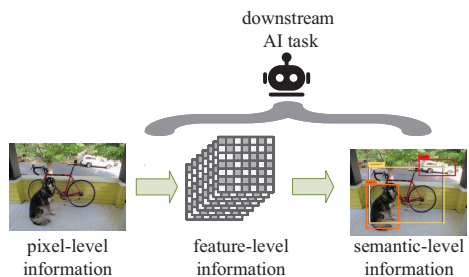}
	\caption{The relationship among pixel-level, feature-level and semantic-level information.}    
    \vspace{-3ex}
	\label{pixel-feature-semantic}
\end{figure}

At the same time, 6G (sixth generation) puts forward the vision of smart interconnection of everything. On the one hand, efficient coding methods are required. On the other hand, the deep integration of communication and artificial intelligence (AI) is required, and research on semantic communication for intelligent tasks \cite{qin2021semantic} has become a trend. Therefore, deep JSCC considering intelligent tasks has gained extensive attention. Deep JSCC methods for image data can be divided into two categories: One is task-oriented without reconstructing the image. The encoder only encodes the semantic information related to the intelligent task, and the decoder directly performs the intelligent task \cite{jankowski2020wireless,liuchuanhong,wang2021deep,yang2021semantic}. This type of work jointly considers communication and intelligence tasks, and explores semantic information extraction for different intelligence tasks. Specifically, Gunduz et al. \cite{jankowski2020wireless} extracted semantic information for image retrieval tasks. Liu et al. \cite{liuchuanhong} extracted semantic information for classification tasks using feature clipping. Wang et al. \cite{wang2021deep} extracted semantic information for various tasks, considering both detection and segmentation tasks. However, this type of work can not apply to the applications that require image reconstruction. The other is that the receiver completes the image reconstruction. The encoder extracts the global semantic information, and the decoder reconstructs the image according to the received semantic information \cite{bourtsoulatze2019deep,kurka2020deep,yang2022deep,xu2021wireless}. Specifically, Gunduz et al. \cite{bourtsoulatze2019deep} designed deep JSCC for image reconstruction, which can surpass SSCC methods in terms of image clarity. Kurka et al. \cite{kurka2020deep} designed the deep JSCC with channel signal feedback. Yang et al. \cite{yang2022deep} designed a rate-adaptive deep JSCC. Xu et al. \cite{xu2021wireless} designed the deep JSCC based on attention, which can flexibly adapt to different SNR (signal-to-noise ratio) conditions. However, all of the above deep JSCC methods for image reconstruction aim at optimizing the visual quality of the image at the receiver. They only focus on the accurate transmission of pixel-level information, while ignoring the semantic information required by downstream AI tasks.

The semantic information of the image refers to final perception results understood by the downstream AI task, which directly affects the performance of the downstream AI task. Fig \ref{pixel-feature-semantic} shows the relationship among pixel-level, feature-level, and semantic-level information. As shown in Fig \ref{pixel-feature-semantic}, feature-level information is the intermediate data of the downstream AI task, while semantic-level information is the object's category and their locations (e.g. dog, bicycle, car). Semantic-level information is further extracted from pixel-level information and feature-level information, and clear pixel-level information is likely to extract correct semantic-level information. However, existing researches show that the relationship among them is not strictly linear, and that means keeping the accurate transmission of pixel-level information does not guarantee the correct understanding of downstream AI tasks \cite{luo2022frequency}. It is worth mentioning that in the field of image compression, some studies have considered content information \cite{li2018learning}, downstream AI task's performance \cite{patwa2020semantic} and feature-level information \cite{yang2020discernible}. However, on the one hand, these studies only consider the source coding, and in the actual communication process, channel conditions and channel coding need to be considered. On the other hand, these studies do not directly utilize semantic-level information, and can not guarantee consistency of the semantic-level information during transmission. In summary, neither the existing deep JSCC method for image reconstruction nor the image compression method considers communication tasks and the semantic-level information of downstream AI tasks jointly. 

To solve the above problems, a semantic importance based deep joint source-channel coding (SD-JSCC) method is proposed for wireless image transmission. By designing the semantic loss function, the image at the receiver can retain the semantic information needed by the downstream AI task, so as to be understood correctly.
For downstream AI tasks, image's feature-level information provides different degrees of contribution to its semantic-level information. 
Based on this, the main contributions are as follows:
\begin{itemize}
  \item We design the gradient-based \cite{selvaraju2017grad} semantic importance weight module to calculate semantic weights.
  This module takes the perception results of the downstream AI task as input and uses its gradient with respect to feature maps to represent the contribution of feature maps to the perception results.
  \item We design the semantic loss function calculation module to calculate the difference of semantic-level information between images at the transmitter and the receiver. We use the difference of weighted features by semantic weights as semantic-level loss function.
  \item  We use the semantic loss function to train the deep joint source-channel codec network in an end-to-end manner, and obtain significant improvement on tasks' performance.
\end{itemize}

\section{System Model and Problem Description}

\subsection{System Model}
\begin{figure}[t]
	\centering
	\includegraphics[width=0.9\linewidth]{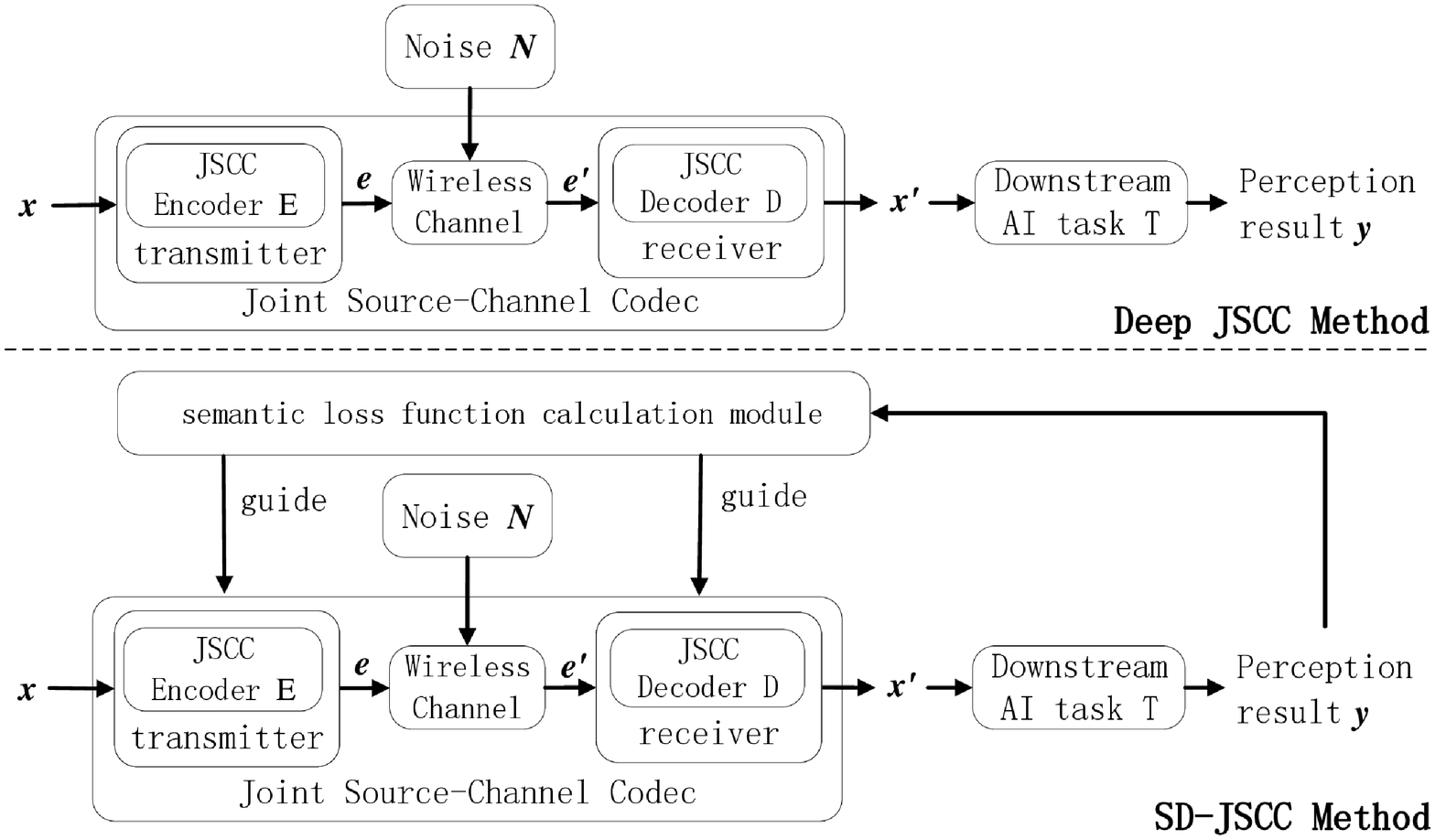}
	\caption{JSCC system model diagram considering AI tasks. Top: deep JSCC method. Bottom: SD-JSCC method.}	\vspace{-4ex}
	\label{system model}
\end{figure}
Fig \ref{system model} shows an end-to-end communication system for wireless image transmission considering the downstream AI task. An input image $\textbf{\textit{x}} \in \mathcal{R}^{\textit{n}}$ of dimension $\textit{n}$ is to be transmitted, where $\mathcal{R}$ denotes the set of real numbers, and the transmitter maps the input image $\textbf{\textit{x}}$ into a complex-valued symbolic vector $\textbf{\textit{e}}$ after the JSCC encoder $\mathrm{E}$, which can be expressed as:
\begin{equation}
\textbf{\textit{e}}=\mathrm{E}\left(\textbf{\textit{x}}, \theta_{1}\right) \in \mathcal{C}^{\textit{s}},
\end{equation}
where $\textit{s}$ denotes the dimension of $\textbf{\textit{e}}$, $\mathcal{C}$ denotes the set of complex numbers, and $\theta_{1}$ denotes the parameters of the JSCC encoder $\mathrm{E}$. The symbol vector $\textbf{\textit{e}}$ after encoding is transmitted over a noisy AWGN channel, which can be expressed as:
\begin{equation}
\textbf{\textit{e}}^{\prime}=\textbf{\textit{e}}+\textbf{\textit{N}} \in \mathcal{C}^{\textit{s}},
\end{equation}
where $\textbf{\textit{N}} \in \mathcal{C}^{\textit{s}}$ denotes the noise of the channel. The noise is obtained by sampling from $\mathbb{C N}\left(0, \sigma^{2} \textbf{\textit{I}}\right)$, where $\sigma^{2}$ denotes the noise power, and $\mathbb{C N}(\cdot, \cdot)$ is a complex Gaussian distribution. The receiver map $\textbf{\textit{e}}^{\prime}$ to the reconstructed image $\textbf{\textit{x}}^{\prime}$ through the JSCC decoder $\mathrm{D}$, which can be expressed as:
\begin{equation}
\textbf{\textit{x}}^{\prime}=\mathrm{D}\left(\textbf{\textit{e}}^{\prime}, \theta_{2}\right)=\mathrm{D}\left(\mathrm{E}\left(\textbf{\textit{x}}, \theta_{1}\right)+\textbf{\textit{N}}, \theta_{2}\right),
\end{equation}
where the reconstructed image $\textbf{\textit{x}}^{\prime} \in \mathcal{R}^{\textit{n}}$ is an estimate of the original image $\textbf{\textit{x}}$ and $\theta_{2}$ is the parameter of the JSCC decoder $\mathrm{D}$. Then, the reconstructed image $\textbf{\textit{x}}^{\prime}$ is passed through the downstream AI task $\mathrm{T}$, and perception results are obtained, which can be expressed as:
\begin{equation}\label{eq4}
\textbf{\textit{y}}=\mathrm{T}\left(\textbf{\textit{x}}^{\prime}\right),
\end{equation}
where $\textbf{\textit{y}}=\left[\textit{y}^{1}, \textit{y}^{2}, \ldots, \textit{y}^{C}\right]$, $\textit{y}^{c}(c \in\{1, \ldots, C\})$ denotes the $\textit{c}$-th perception result and $C$ is the total number of perception results.
\subsection{Problem Description}
The existing deep JSCC methods \cite{bourtsoulatze2019deep,kurka2020deep,yang2022deep,xu2021wireless} use the pixel-level difference between $\textbf{\textit{x}}^{\prime}$ and  $\textbf{\textit{x}}$ as the loss function to train the joint source-channel codec network, which can be expressed as:
\begin{equation}
\mathcal{L}_{\text {deep JSCC }}=\textit{d}\left(\textbf{\textit{x}}, \textbf{\textit{x}}^{\prime}\right)=\left\|\textbf{\textit{x}}-\textbf{\textit{x}}^{\prime}\right\|^{2}.
\end{equation}
$\mathcal{L}_{\text {deep JSCC }}$ enables $\textbf{\textit{x}}^{\prime}$ to obtain a clear visual quality that is close to the original image $\textbf{\textit{x}}$. This approach only maintains the pixel-level consistency during image transmission, without considering the perception results of the downstream AI task.

In addition, there are some deep learning-based source coding methods [14] that use feature-level differences to train source coding networks, which can be expressed as:
\begin{equation}
\mathcal{L}_{\text {source }}=\textit{d}(\mathrm{F}(\textbf{\textit{x}}), \mathrm{F}(\tilde{\textbf{\textit{x}}}))=\|\mathrm{F}(\textbf{\textit{x}})-\mathrm{F}(\tilde{\textbf{\textit{x}}})\|^{2},
\end{equation}
where $\mathrm{F}(\cdot)$ denotes the feature-level information, and $\tilde{\textbf{\textit{x}}}$ denotes the reconstructed image obtained only by source coding and decoding. On the one hand, this is a source coding approach, which does not consider channel coding and the actual communication process. The network can not learn the features of the channel and can not learn how to resist the channel noise. On the other hand, $\mathcal{L}_{\text {source }}$ only aims at the consistency of feature-level information. However, feature-level information has different semantic importance for downstream AI tasks.

Therefore, in order to jointly consider the performance of communication tasks and downstream AI tasks, it is necessary to consider the semantic-level information and perception results of the image at the receiver. Studying JSCC that preserves the semantic information of downstream AI tasks during image transmission is meaningful.

\section{SD-JSCC Method}
\begin{figure*}[t]
	\centering
	\includegraphics[width=0.8\linewidth]{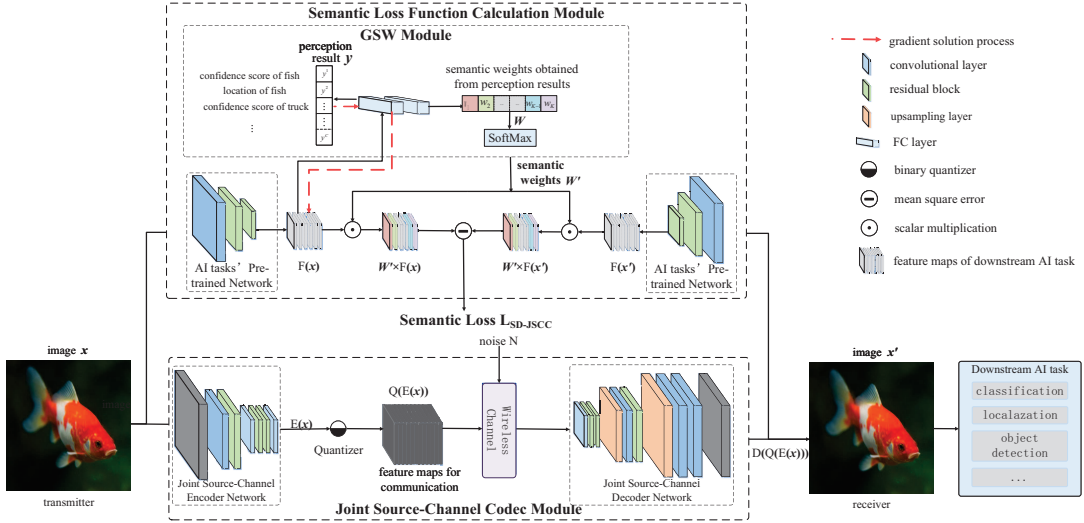}
	\caption{SD-JSCC network architecture diagram}    
	\vspace{-3ex}
	\label{architecture}
\end{figure*}
As shown in Fig \ref{system model}, in order to preserve the semantic-level information for the downstream AI task, the SD-JSCC method uses the perception results of the downstream AI task to calculate the semantic loss function and guide the joint source-channel codec module. As shown in Fig \ref{architecture}, SD-JSCC's architecture consists of two major modules, namely, the semantic loss function calculation module and the joint source-channel codec module. First, we propose the semantic loss function calculation module, which can characterize the semantic-level distortion between $\textbf{\textit{x}}^{\prime}$ and $\textbf{\textit{x}}$. Then, the semantic loss function is used to train the joint source-channel codec module.

The semantic loss function calculation module is the core of the SD-JSCC method. The feature-level information of the AI task provides different degree of contribution to its semantic-level information. Therefore, the semantic loss function is constructed using semantic weights to weight feature-level information, so that can preserve the semantic-level information of the downstream AI task. Semantic weights are obtained through the gradient-based semantic importance weight (GSW) module, which draws on ideas related to the field of neural network interpretability [16]. It is acknowledged that using gradient to measure the semantic importance for explaining the neural network learning process.

Specifically, first, we use the GSW module to calculate the semantic weights $\textbf{\textit{W}}^{\prime}$. Then, we calculate the semantic loss function using $\textbf{\textit{W}}^{\prime}$ to weight the feature-level information. Finally, we use the semantic loss function $\mathcal{L}_{\text {deep JSCC }}$ to train the joint source-channel codec module.
\subsection{GSW Module}

This section designs the GSW module for semantic weights $\textbf{\textit{W}}^{\prime}$ calculation and gives the formula representation of $\textbf{\textit{W}}^{\prime}$.

As shown in Fig \ref{architecture}, the GSW module is designed to calculate the semantic weights to quantify the importance of the feature map to downstream AI task's semantic information. The GSW module calculate the gradient of perception results with respect to the feature map, which can naturally represent the contribution of the feature map to the semantic information understood by the downstream AI task.

First, the network of downstream AI tasks is pre-trained to extract feature-level information and provide input for the  semantic weights calculation, which can be expressed as:
\begin{equation}
\mathrm{F}\left(\theta_{0}, \textbf{\textit{x}}\right)=\left\{\textbf{\textit{f}}_{1}, \textbf{\textit{f}}_{2}, \ldots, \textbf{\textit{f}}_{\mathrm{K}}\right\} \in \mathcal{R}^{\textit{K} \times \textit{M} \times \textit{N}},
\end{equation}
where $\textbf{\textit{f}}_{\textit{k}}(\textit{k} \in\{1,2, \ldots, \textit{K}\})$ denotes the $\textit{k}$-th feature map, $\theta_{0}$ denotes the fixed parameters of the feature extraction network, and $\textit{M}$, $\textit{N}$, $\textit{K}$  denote the width, height and number of feature maps, respectively. 

Next, the gradient of the perceptual results with respect to the feature-level information is used to quantify the semantic importance. The gradient of the $\textit{c}$-th perception result $\textit{y}^\textit{c}$ to the $\textit{k}$-th feature map $\textbf{\textit{f}}_\textit{k}$ can be denoted as $\frac{\partial \textit{y}^{\textit{c}}}{\partial \textit{f}_{\textit{k}}}$. The average value over the width $\textit{M}$ and height $\textit{N}$ dimensions can be expressed as:
\begin{equation}
\textit{w}_{\textit{k}}^{\textit{c}}=\frac{1}{\textit{M} \times \textit{N}} \sum_{\textit{m}=1}^{\textit{M}} \sum_{\textit{n}=1}^{\textit{N}} \frac{\partial \textit{y}^{\textit{c}}}{\partial \textbf{\textit{f}}_{\textit{k}}},
\end{equation}
where $\textbf{\textit{f}}_{\textit{k}} \in \mathcal{R}^{\textit{M} \times \textit{N}}$. To represent the comprehensive influence of feature map $\textbf{\textit{f}}_{\textit{k}}$ on all perception results $\textbf{\textit{y}}$, the average value of $\textit{w}^\textit{c}_\textit{k}$ over all perception results is calculated, which can be expressed as:
\begin{equation}
\textit{w}_{\textit{k}}=\frac{1}{\textit{C}} \sum_{\textit{c}=1}^{\textit{C}} \textit{w}_{\textit{k}}^{\textit{c}}.
\end{equation}
Then, the semantic weight vector $\textbf{\textit{W}}=\left\{\textit{w}_{1}, \textit{w}_{2}, \ldots, \textit{w}_{\textit{K}}\right\} \in \mathcal{R}^{\textit{K}}$ is obtained, and $\textit{w}_{\textit{k}}(\textit{k} \in\{1,2, \ldots, \textit{K}\})$ represents the semantic importance of the $\textit{k}$-th feature map $\textbf{\textit{f}}_{\textit{k}}$.

However, the value of $\textbf{\textit{W}}$ is too small to be directly used in loss function, since it may cause the slow convergence. Therefore, we utilize the parameters $\tau$ and $\textit{r}$, and map $\textbf{\textit{W}}$ to $\textbf{\textit{W}}^{\prime}$ by:
\begin{equation}
\textbf{\textit{W}}^{\prime}=\textit{r} \times \operatorname{SoftMax}(\tau \times \textit{W}),
\end{equation}
where $\tau$ is a temperature hyper-parameter and $\textit{r}$ is a constant. The semantic weights $\textbf{\textit{W}}^{\prime}$ are appropriate for semantic-level loss function. The temperature hyper-parameter $\tau$ can control the distribution of the semantic weights $\textbf{\textit{W}}^{\prime}$. The constant $\textit{r}$ can make the final semantic-level loss value at a
reasonable magnitude, which will not affect gradient updates.

Note that since the design of the GSW module, the semantic weights are specific to the downstream AI task, and therefore SD-JSCC is also task-specific.

\subsection{Semantic Loss Function}
We introduce this section to formulate semantic-level loss
function $\mathcal{L}_{\text {SD-JSCC }}$ for wireless image transmission.

The feature-level information of the downstream AI task at the transmitter and receiver are extracted respectively by:
\begin{equation}
\mathrm{F}\left(\theta_{0}, \textbf{\textit{x}}_{\mathrm{b}}\right)=\left\{\textbf{\textit{f}}_{1}^{\textit{b}}, \textbf{\textit{f}}_{2}^{\textit{b}}, \ldots, \textbf{\textit{f}}_{\textit{K}}^{\textit{b}}\right\},
\end{equation}
\begin{equation}
\mathrm{F}\left(\theta_{0}, \textbf{\textit{x}}_{\mathrm{b}}^{\prime}\right)=\left\{\textbf{\textit{f}}_{1}^{\textit{b}^{\prime}}, \textbf{\textit{f}}_{2}^{\textit{b}^{\prime}}, \ldots, \textbf{\textit{f}}_{\textit{K}}^{\textit{b}^{\prime}}\right\},
\end{equation}
where $\textbf{\textit{x}}_{\textit{b}}$ is the $\textit{b}$-th image at the transmitter, and $\textbf{\textit{x}}_{\textit{b}}^{\prime}$ is the $\textit{b}$-th image at the receiver. $\textbf{\textit{f}}_{\textit{k}}^{\textit{b}}$ is the $\textit{k}$-th feature map of image $\textbf{\textit{x}}_{\textit{b}}$ , and $\textbf{\textit{f}}_{\textit{k}}^{\textit{b}^{\prime}}$ is the $\textit{k}$-th feature map of image $\textbf{\textit{x}}_{\textit{b}}^{\prime}$, $\textit{k} \in\{1,2, \ldots, \textit{K}\}$.

Then, the semantic loss function is calculated using the semantic weights $\textbf{\textit{W}}^{\prime}=\left\{\textit{w}_{1}^{\prime}, \textit{w}_{2}^{\prime}, \ldots, \textit{w}_{\textit{K}}^{\prime}\right\} \in \mathcal{R}^{\textit{K}}$ derived from the GSW module to weight feature-level information, which can be expressed as:
\begin{equation}
\mathcal{L}_{\text {SD-JSCC }}\left(\theta_{1}, \theta_{2}\right)=\frac{1}{\textit{B}} \sum_{\textit{b}=1}^{\textit{B}} \sum_{\textit{k}=1}^{\textit{K}} \textit{w}_{\textit{k}}^{\prime} \times\left\|\textbf{\textit{f}}_{\textit{k}}^{\textit{b}^{\prime}}-\textbf{\textit{f}}_{\textit{k}}^{\textit{b}}\right\|^{2},
\end{equation}
where $\textit{B}$  (batch size) is the number of images per iteration.

\subsection{Training Joint Source-Channel Codec Module}
The architecture of the joint source-channel codec network is shown in Fig \ref{architecture}. The encoder consists of convolution layers and residual blocks, and the decoder consists of upsampling layers, convolution layers and residual blocks. First, the image is encoded by the joint source-channel encoder. Then the noise in the wireless channel is simulated. Finally, we use the joint source-channel decoder to decode and recover the image. For convenience, $\textbf{\textit{e}}=\mathrm{E}\left(\theta_{1}, \textbf{\textit{x}}\right)$, $\textbf{\textit{q}}=\mathrm{Q}\left(\mathrm{E}\left(\theta_{1}, \textbf{\textit{x}}\right)\right)$, $\textbf{\textit{d}}=\mathrm{D}\left(\mathrm{Q}\left(\mathrm{E}\left(\theta_{1}, \textbf{\textit{x}}\right)\right)+\textbf{\textit{N}}, \theta_{2}\right)$ are used to denote the output of the encoder, quantizer, and the decoder, respectively. $\mathrm{E}(\cdot)$ is the joint source-channel encoder network for encoding the image into a hidden layer representation with parameters $\theta_{1}$. $\mathrm{Q}(\cdot)$ is the quantizer. $\mathrm{D}(\cdot)$ is the joint source-channel decoder network from recovering the image from the hidden layer representation with parameters $\theta_{2}$. Note the quantizer is designed to reduce the transmitter's cost \cite{xie2020lite}, and can map each element $\textit{e}_{\textit{i}, \textit{j}}$ of the encoder output $\textbf{\textit{e}}$ to 0 or 1, which can be expressed as:
\begin{equation}
\mathrm{Q}\left(\textit{e}_{\textit{i}, \textit{j}}\right)=\left\{\begin{array}{l}
1, \enspace \textit{e}_{\textit{i}, \textit{j}}>0.5 \\
0, \enspace \textit{e}_{\textit{i}, \textit{j}} \leq 0.5
\end{array}\right.
\end{equation}
This end-to-end process is trained using the semantic loss function so that the semantic information beneficial to the downstream AI task is retained.
\begin{algorithm}[t]
	\caption{SD-JSCC method}
	\label{algorithm1}
	\textbf{Input}: An image dataset $\left\{{{\boldsymbol{x}}^1},...,{{\boldsymbol{x}}^n}\right\}$ with $n$ images.\\
	\textbf{Parameter}: Encoder parameter ${\theta _1}$, decoder parameter ${\theta _2}$.\\
	\textbf{Output}: Parameters ${\theta _1}$ and ${\theta _2}$, images $\textbf{\textit{x}}^{\prime}$ at receiver, hidden feature maps $\textbf{\textit{q}}$.
	\begin{algorithmic}[1] 
		\STATE Pre-train the downstream AI task's network with parameter ${\theta _0}$, and fixed ${\theta _0}$ in the following operation.
		\STATE Obtain the semantic weights $\textbf{\textit{W}}^{\prime}$ using GSW.
		\STATE Initialize encoder and decoder's parameter ${\theta _1}$, ${\theta _2}$.
		\WHILE{ not converged }
		\STATE Image Encoding: $\textbf{\textit{e}} \leftarrow E({\theta _1},\textbf{\textit{x}})$.
		\STATE Quantization: $\textbf{\textit{q}} \leftarrow  Q(\textbf{\textit{e}})$.
		\STATE Noise channel and image decoding: $\textbf{\textit{x}}^{\prime} = \textbf{\textit{d}} \leftarrow D({\theta _2},\textbf{\textit{q + N}})$.
		\STATE Calculate feature maps according to equation (11), (12).
		\STATE Calculate the semantic loss function $\mathcal{L}_{\text {SD-JSCC}}$.
		\STATE Update ${\theta _1}$ and ${\theta _2}$ according to $\mathcal{L}_{\text {SD-JSCC}}$.
		\ENDWHILE
		\STATE \textbf{return} The optimal model.
	\end{algorithmic}
\end{algorithm}

We summarize the steps of SD-JSCC as Algorithm 1. In step
1, we pre-train the downstream AI task's network. In step 2, we
obtain the semantic weights. In step 3-9, we compute the semantic loss function $\mathcal{L}_{\text {SD-JSCC}}$. In step 10, we use  $\mathcal{L}_{\text {SD-JSCC}}$ to train the joint source-channel codec network.

\section{Experiment}

\subsection{Evaluation Metrics and Comparison Methods}
This section evaluates the distortion of semantic information, the distortion of pixel information, and compression degree, respectively. We use the downstream AI task's performance at the receiver to evaluate the distortion of semantic information. For classification tasks, we use accuracy (ACC) and F1-score. For object detection tasks, we use mean Average Precision (mAP). To evaluate the distortion of pixel information, we use the peak signal-to-noise ratio (PSNR) and the structural similarity index measure (SSIM). To evaluate the degree of compression, we use the compression rate (bit per pixel, bpp).
The traditional SSCC scheme and the existing mainstream JSCC scheme are compared respectively.
For the SSCC scheme, JPEG, WebP and BPG are used as the source coding method. LDPC is used as the channel encoding mode, and LDPC codes is (1458, 1944), corresponding to rate 3/4. 16-QAM is used as the modulation method. 
For the JSCC scheme, the deep JSCC method proposed by Gunduz et al \cite{bourtsoulatze2019deep} is used for comparison. For fair comparison, SD-JSCC uses the same experimental conditions with deep JSCC except for the loss function, including the encoder-decoder network structure and quantization method.
\vspace{-1ex}
\subsection{ Implementation Details}
For classification tasks, we adpot STL and a split of ImageNet as datasets. For object detection task, we adpot Pascal VOC dataset.
For channel, We consider AWGN.
In order to reduce training costs and promote extensibility, the SD-JSCC model is trained using a two-stage approach. In the first stage, the deep JSCC model is pre-trained with 3-4$\times$10$^{5}$ steps using a batch size of 32 and a learning rate of 1$\times$10$^{-5}$. In the second stage, parameters obtained in the first stage are loaded, then we finetune the network using the SD-JSCC method with 2-3$\times$10$^{4}$ steps, where the batch size is 32 and the learning rate is 1$\times$10$^{-5}$.

The original image in the subsequent experiment results refers to the original image in the dataset. The compression rate of the original image is 15.38bpp and 4.61bpp on STL and ImageNet, respectively. For SSCC methods (e.g. JPEG, WebP, BPG), the compression rate can not be assigned directly and an approximate bpp value is obtained. For the JSCC method, the exact compression rate can be assigned by changing the dimension of the encoder's output $\textbf{\textit{e}}$.
\subsection{Performance Evaluation and Analysis}
\subsubsection{SD-JSCC Overall Performance Assessment}
\begin{figure}[t]
	\centering
	\includegraphics[width=0.6\linewidth]{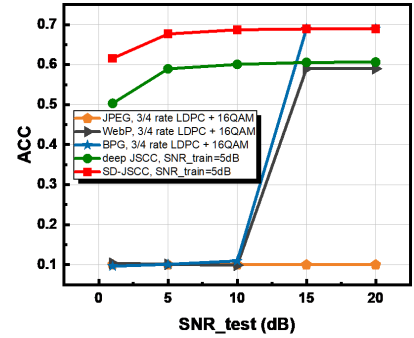}
	\caption{Performance comparison of downstream AI tasks using different methods on STL at 0.25bpp.}    
	\vspace{-3.5ex}
	\label{overall}
\end{figure}

Fig \ref{overall} gives the variation curve of ACC with SNR of different methods, which is done on STL at 0.25bpp. SNR\_train=5dB indicates the SNR at training is 5dB, and SNR\_test indicates the SNR value at testing. As shown in Fig \ref{overall}, SD-JSCC can achieve significant performance improvement under low SNR. For example, ACC is improved by 9.1\% and 57.7\% compared with the deep JSCC method and the SSCC method at 5dB, respectively. This is due to the semantic loss function designed by the SD-JSCC method, which can retain semantic information that is beneficial to downstream AI tasks, and deservedly
can obtain competitive ACC values.
The SSCC method performs well under high SNR, but when the SNR falls below a threshold, ACC decreases sharply, which is called "cliff effect" \cite{bourtsoulatze2019deep}.
JPEG method's ACC has been kept at the lowest level since the image restoration quality of JPEG method is particularly poor at 0.25bpp. JPEG method can recover the image only when the compression rate is above 0.75bpp on STL. WebP and BPG can not assign compression rate directly, and the estimation
value is 0.27bpp and 0.269bpp in practice, respectively. 
Table 1 compares the task performance (ACC, F1-score) and pixel-level metrics (PSNR, SSIM) of the different methods. This experiment is completed on ImageNet at 0.125bpp under good channel conditions. As shown in table 1, ACC and F1-score are more competitive using SD-JSCC. For example, ACC achieves 10.94\% and 2.59\% improvement compared to deep JSCC method and the best SSCC method (BPG), respectively. 
The SD-JSCC approach is designed with a semantic loss function that enables the network focus on semantic consistency during transmission directly, rather than pixel-level consistency. Therefore, SD-JSCC can achieve better task performance although pixel-level metrics are slightly poor. Due to the cliff effect under low SNR, the PSNR and SSIM of SD-JSCC method will exceeding these SSCC method. At the same time, ACC and F1-score are close to each other, which is due to the balanced distribution of categories on dataset. Therefore, it is sufficient to refer to one of them to evaluate the performance of classification task, and subsequently only use ACC for evaluation.

\begin{figure*}[htbp]
	\centering
	\subfigure[carp]{
		\begin{minipage}[t]{0.12\linewidth}
			\includegraphics[width=0.8in]{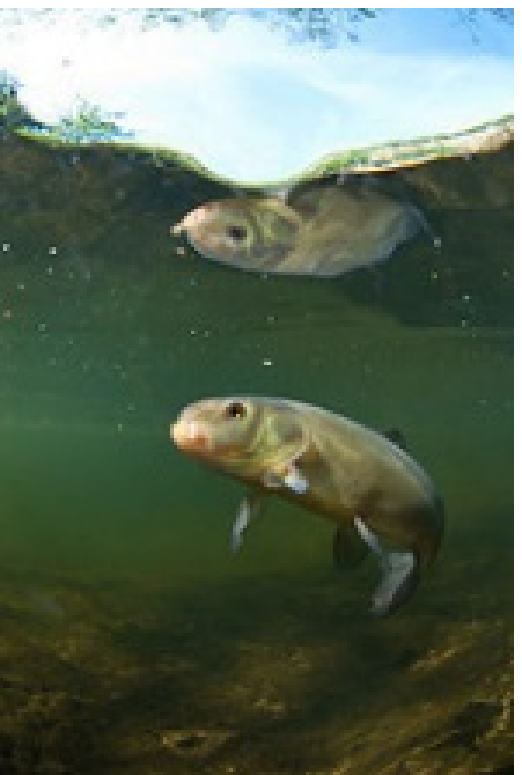}
		\end{minipage}%
	}
	\subfigure[tiger shark 
	\XSolidBrush
	]{
		\begin{minipage}[t]{0.12\linewidth}
			\centering
			\includegraphics[width=0.8in]{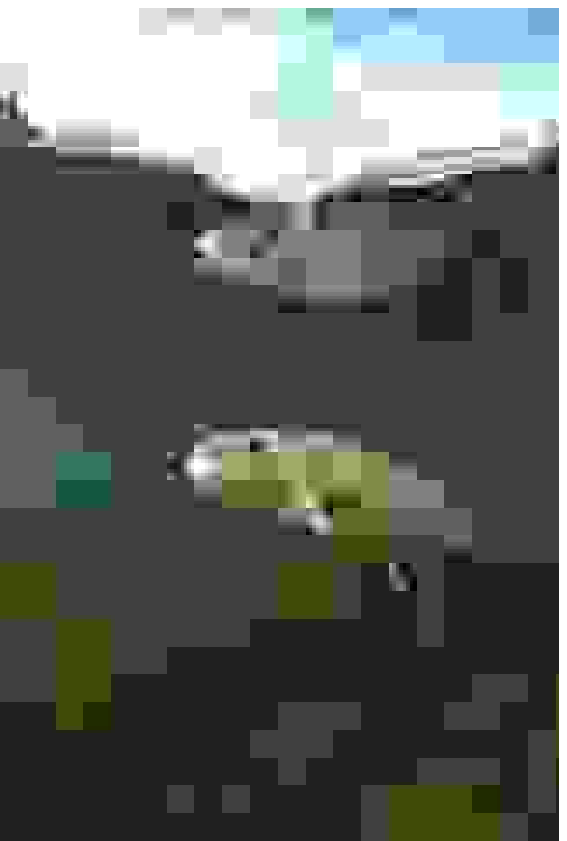}
		\end{minipage}%
	}%
	\subfigure[Stingray
	\XSolidBrush
	]{
		\begin{minipage}[t]{0.12\linewidth}
			\centering
			\includegraphics[width=0.8in]{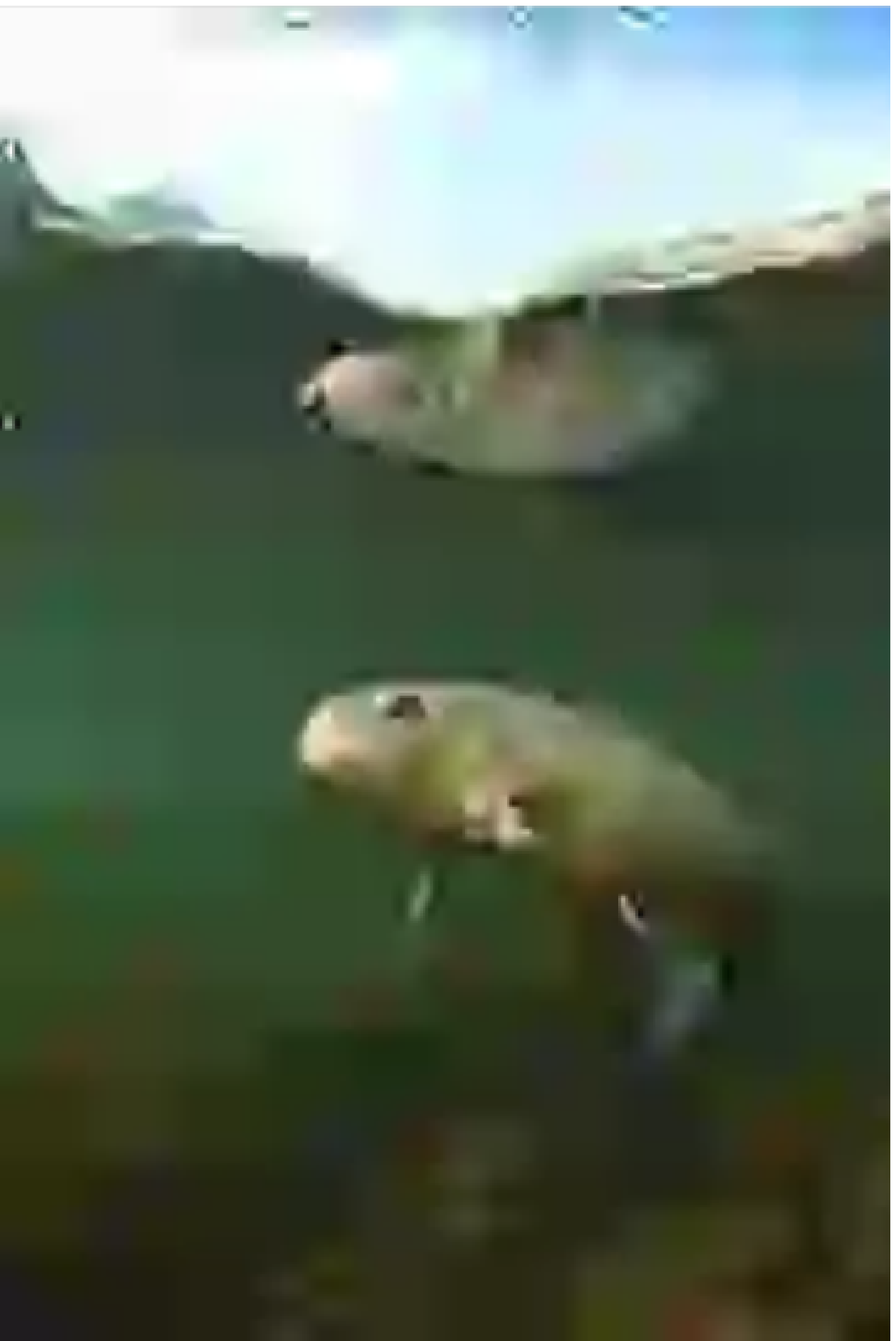}
		\end{minipage}
	}%
	\subfigure[tiger shark
	\XSolidBrush]{
		\begin{minipage}[t]{0.12\linewidth}
			\centering
			\includegraphics[width=0.8in]{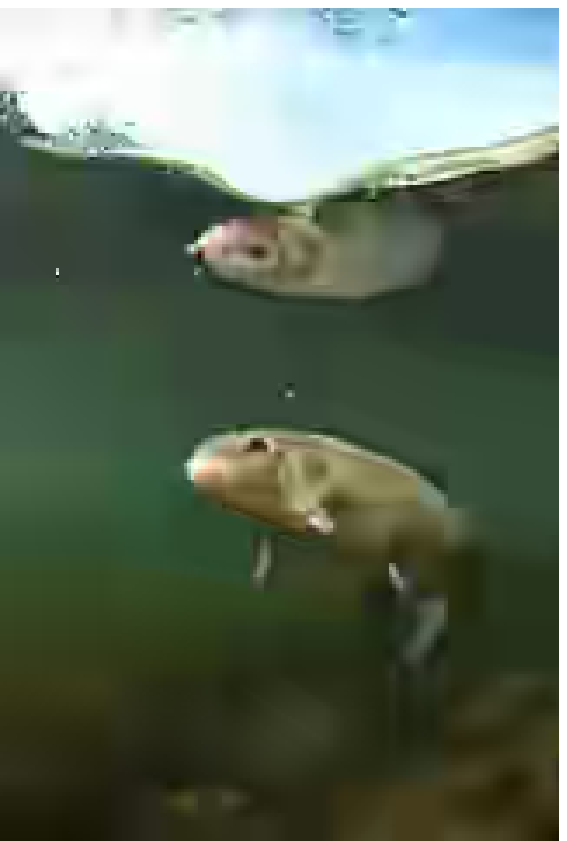}
		\end{minipage}
	}
	\subfigure[electric rays
	\XSolidBrush]{
		\begin{minipage}[t]{0.12\linewidth}
			\centering
			\includegraphics[width=0.8in]{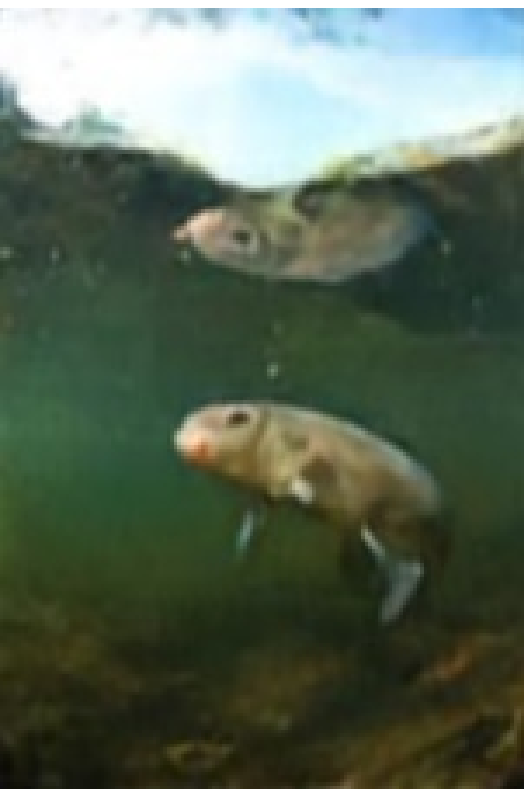}
		\end{minipage}	
	}
	\subfigure[carp \CheckmarkBold]{
		\begin{minipage}[t]{0.12\linewidth}
			\centering
			\includegraphics[width=0.8in]{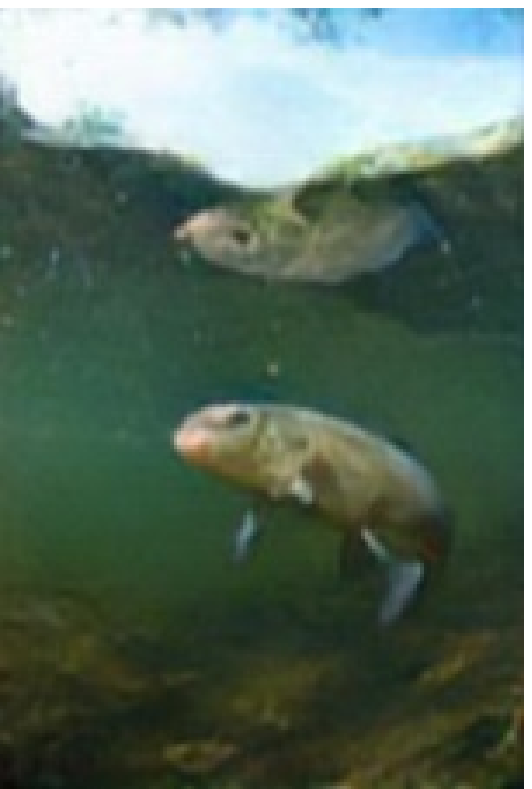}
		\end{minipage}	
	}
	\centering
	\caption{Comparison of images and their perception results of classification task at the receiver. From left to right are original image, reconstruction image using JPEG, WebP, BPG, deep-JSCC, SD-JSCC, respectively.}
	\vspace{-3ex}
	\label{fish}
	
\end{figure*}

Fig \ref{fish} compares the image at the receiver of different methods. As shown in Fig \ref{fish}, there are subtle differences on images, but totally different perception results. Pixel-level consistency cannot guarantee the consistency of downstream AI task's perception results, and some tiny distortions at the pixel level may cause false perception results.

\begin{table}
	\scriptsize
	\centering
	\caption{Comparison of multiple evaluation metrics using different methods on ImageNet at 0.125bpp.}
	\label{table6}
	\setlength{\abovecaptionskip}{-0.2cm}
	\begin{tabular}{llllll}
	\toprule 
		Method & ACC & F1-score & PSNR & SSIM \\
		\midrule
		original & 89.42\%  & 0.8937 & - & 1 \\
		JPEG & 39.78\% & 0.3954 & 22.00 & 0.5299 \\
		WebP & 74.50\% & 0.7448 & 27.83 & 0.6885 \\
		BPG & 79.57\% & 0.7953 & 29.39 & 0.7318 \\
		deep JSCC & 71.22\% & 0.7125 & 28.89 & 0.8403 \\
		SD-JSCC & \pmb{82.16\%} & 0.8214 & 27.02 & 0.7641 \\
		 \bottomrule
	\end{tabular}
\end{table}

\subsubsection{SD-JSCC Ablation Experiment}

\begin{figure}[t]
	\centering
	\includegraphics[width=0.8\linewidth]{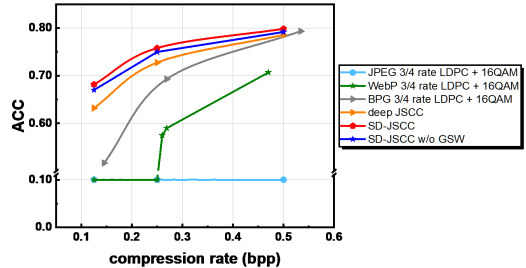}
	\caption{Performance comparison of the downstream AI task using different compression rate on STL.}    
	\label{ablation}
\end{figure}

The core of the SD-JSCC method is the semantic loss function calculation module, and the GSW module is a part of the semantic loss function calculation module. In order to further explore the contribution of the GSW module to the SD-JSCC method, we design an ablation experiment.
We evaluate ACC for wireless image transmission when the GSW module fails, subsequently expressed as SD-JSCC w/o GSW.
Fig \ref{ablation} gives the variation curve of ACC with compression rate on STL dataset with SNR\_train=SNR\_test=20dB.
As shown in Fig \ref{ablation}, compared with SD-JSCC w/o GSW method, SD-JSCC method has higher ACC value under the same compression ratio. Conversely, under the same ACC, that is, the same semantic distortion, the SD-JSCC method can compress more. 
This is because the semantic weights of all feature maps in the SD-JSCC w/o GSW method are equal to 1, and the semantic weights obtained by GSW module can correctly represent the importance degree of feature maps.
The ablation experiments show that the overall performance advantage of the SD-JSCC method consists of two parts in the semantic loss function calculation module, one part is the GSW module, and the other part is the feature extraction network.

\begin{figure}[t]
	\centering
	\includegraphics[width=0.5\linewidth]{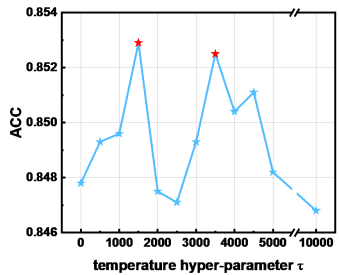}
	\caption{Performance of the downstream AI task under different temperature hyper-parameters on ImageNet at 0.5bpp.}    
	\vspace{-3ex}
	\label{temperature h-p}
\end{figure}

To assess the influence of the temperature hyper-parameter $\tau$, the variation curve of ACC with $\tau$ for the SD-JSCC method is shown in Fig \ref{temperature h-p}, which is done on ImageNet at 0.5bpp with good channel conditions.
As shown in Fig \ref{temperature h-p}, with $\tau$ increases, the trend of ACC is first up, then down, then up and down again. This is because the hyper-parameter $\tau$ can control the dispersion degree of semantic weights distribution, and extremely concentrated ($\tau$ =1 and $\tau$ =10000) or decentralized ( $\tau$ =2000) semantic weights can degrade the performance.
Excessively concentrated semantic weights make the importance degree of different feature maps almost the same, so SD-JSCC degenerates to SD-JSCC w/o GSW.
Excessively decentralized semantic weights are equivalent to deleting some feature maps, resulting in the missing of some feature information. Only appropriate degree of dispersion can obtain good performance improvement. The optimal value of $\tau$ varies with model structure and dataset.

\subsubsection{SD-JSCC Robustness Analysis}
\begin{figure}[t]
	\centering
	\includegraphics[width=0.5\linewidth]{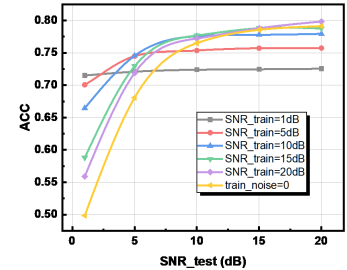}
	\caption{Performance of the downstream AI task with different SNR using SD-JSCC method on STL at 0.5bpp.}    
	\vspace{-0.3ex}
	\label{robust}
\end{figure}

In order to explore the robustness of SD-JSCC method to channel conditions, Fig \ref{robust} shows the variation curve of ACC with SNR, which is performed at 0.5bpp on STL dataset.
Each curve represents SD-JSCC method's acc optimised under  SNR\_train and deployed under SNR\_test.
It can be seen from Fig \ref{robust} that SD-JSCC is robust to channel quality fluctuation.
When SNR\_test is worse than SNR\_train, the SD-JSCC method does not suffer from cliff effect of the SSCC method, and shows a gradual performance degradation as the channel conditions deteriorate.
Similarly, as SNR\_test increases above SNR\_train, the quality of the image at the receiver gradually improves, and when SNR\_test increases to a certain value, the performance eventually saturates.
This is because the neural network maps similar image features to nearby points in the original image feature space. Thus, under low SNR, the decoder can still obtain a reconstruction of the original image, and there is no cliff effect.

\subsubsection{ SD-JSCC Generalizability Analysis}
To investigate the generalisation performance of the SD-JSCC method on different tasks, we chose object detection task as the downstream AI task.
Table 2 compares mAP of different methods under different SNR, which on Pascal VOC0712 dataset at 0.25bpp using RFBNet-300. 
As shown in Table 2, mAP of SD-JSCC method still exceeds deep JSCC method and SSCC method under low SNR.
This is because the semantic information learned by the SD-JSCC method on the classification task is still effective for object detection task.
BPG method outperforms SD-JSCC method at high SNR, since SSCC method has more compression space and better performance for high image resolutions. To further improve the performance of object detection task, the semantic-level information of object detection task can be used to train the joint source-channel codec network in the future.

\begin{table}
	\scriptsize
	\centering
	\caption{Task performance comparison of object detection}
	\label{table6}
	\setlength{\abovecaptionskip}{-0.2cm}
	\begin{tabular}{lllllll}
		\toprule 
		SNR & JPEG & WebP & BPG & deep JSCC & SD-JSCC \\
		\midrule
		1 dB & - & -  & - & 45.79\% & 51.11\% \\
		5 dB & - & -  & - & 52.70\% & 53.70\% \\
		10 dB & - & 0.20\% & 0.48\% & 58.99\% & 62.53\% \\
		11 dB & - & 23.02\% & 41.62\% & 59.25\% & 62.89\% \\
		15 dB & 35.44\% & 65.94\% & 73.47\% & 64.91\% & 66.42\% \\
		20 dB & 35.44\% & 65.94\% & 73.50\% & 66.63\% & 68.34\% \\
		 \bottomrule
	\end{tabular}
	\vspace{-0.5cm}
\end{table}

\section{conclusion}

In order to improve the downstream AI task's performance of image at the receiver in the wireless image transmission system, we propose SD-JSCC module, and give its network structure and algorithm. 
SD-JSCC method extracts semantic weights by GSW module and designs semantic loss function using downstream AI task's perception results. Therefore, the image at the receiver retains the semantic information needed by the downstream AI task, so as to be understood correctly. The experimental results show that SD-JSCC method significantly improves the performance of downstream AI tasks, and can be used for a wide range of AI tasks. 
The drawback of SD-JSCC is that it increases training overhead, which can be mitigated by fine-tuning on the pre-trained model. However, there is no additional testing overhead.
In the future, we will consider other downstream AI tasks, and simplify the network to facilitate deployment.



\bibliographystyle{unsrt}
\bibliography{conference_101719}

\end{document}